\documentclass[prb,twocolumn,showpacs,superscriptaddress]{revtex4-1}
\usepackage{amsmath}
\usepackage{amssymb}
\usepackage{graphicx}
\usepackage{longtable}%
\usepackage{dcolumn}
\usepackage{bm}
\usepackage{color}
\begin{document}
\title{Low-temperature phase diagram of  Fe$_{1+y}$Te \\}
\author{Cevriye Koz}
\affiliation{Max Planck Institute for Chemical Physics of Solids,
N\"othnitzer Stra\ss e 40, 01187 Dresden, Germany}
\author{Sahana R\"o{\ss}ler}
\email{roessler@cpfs.mpg.de}
\affiliation{Max Planck Institute for Chemical Physics of Solids,
N\"othnitzer Stra\ss e 40, 01187 Dresden, Germany}
\author{Alexander A. Tsirlin}
\affiliation{Max Planck Institute for Chemical Physics of Solids,
N\"othnitzer Stra\ss e 40, 01187 Dresden, Germany}
\affiliation{National Institute of Chemical Physics and Biophysics, 12618 Tallinn, Estonia }
\author{Steffen Wirth}
\affiliation{Max Planck Institute for Chemical Physics of Solids,
N\"othnitzer Stra\ss e 40, 01187 Dresden, Germany}
\author{Ulrich~Schwarz}
\email{schwarz@cpfs.mpg.de}
\affiliation{Max Planck Institute for Chemical Physics of Solids,
N\"othnitzer Stra\ss e 40, 01187 Dresden, Germany}
\date{\today}
%
%
%
\begin{abstract}
We used low-temperature synchrotron x-ray diffraction to investigate the structural phase transitions of Fe$_{1+y}$Te in the vicinity of a tricitical point in the phase diagram. Detailed analysis of the powder diffraction patterns and temperature dependence of the peak-widths in Fe$_{1+y}$Te showed that two-step structural and magnetic phase transitions occur within the compositional range 0.11 $\leq  y  \leq $ 0.13. The phase transitions are sluggish indicating a strong competition between the orthorhombic and the  monoclinic phases. We combine high-resolution diffraction experiments with specific heat, resistivity, and magnetization measurements and present a revised  temperature-composition phase diagram for Fe$_{1+y}$Te.
\end{abstract}
\maketitle
\section{Introduction}

Iron chalcogenides, Fe$_{1+y}$(Te,Se) are promising candidates to understand the mechanism of  superconductivity in the family of Fe-based superconductors owing to their archetypical binary atomic pattern. The tetragonal PbO-type Fe$_{1+y}$Se with a superconducting transition temperature $T_{c}$ = 8~K is the simplest member of Fe-based superconductors because of its structure and chemical composition. \cite{Hsu2008} The structure comprises stacks of edge-sharing FeSe$_{4}$ tetrahedra, which form layers orthogonal to the $c-$axis. The homogeneity range of tetragonal Fe$_{1+y}$Se is very narrow.  The compound is nearly stoichiometric, and minute change in the composition controls the physical and low temperature structural properties. For example, Fe$_{1.01}$Se is superconducting and the crystal structure transforms from  a tetragonal  ($P4/nmm$) to an orthorhombic ($Cmma$) phase at around 90~K, whereas non-superconducting Fe$_{1.03}$Se does not exhibit this structural transition. \cite{McQueen2009} 
The  $T_{c}$ of Fe$_{1+y}$Se can be enhanced up to 37~K by applying external pressure of $7-9$ GPa,  \cite{Mizuguchi2008, Medvedev2009, Margadonna2009}or up to 15~K by about 50 \% substitution of Te at ambient pressure. \cite{Yeh2008,Fang2008,Ros2010} The bulk superconductivity disappears with higher Te substitution and the end member, Fe$_{1+y}$Te, is non-superconducting. 

Fe$_{1+y}$Te with an analogous crystal structure to Fe$_{1+y}$Se occurs only in the presence of excess Fe, which is situated  in the interstitial $2c$ crystallographic sites within the chalcogenide planes. \cite{Bao2009} Instead of superconductivity, tetragonal Fe$_{1+y}$Te shows a complex interplay of magnetic and structural phase transitions in dependence of the excess amount of Fe. \cite{Bao2009, Li2009, Hu2009, Rodriguez 2011, Zaliznyak2012, Ros2011}
A simultaneous first-order magnetic and structural transition from the tetragonal paramagnetic to the monoclinic ($P2_{1}/m$) commensurate antiferromagnetic phase is observed at $T = $~69~K in Fe$_{1.06}$Te. The first-order transition temperature systematically decreases down to 57 K with an increase in $y$ from 0.06 to 0.11. For $ y > 0.11$, two transitions are observed: in the specific case of $y=$~0.13, a continuous transition at 57 K and a first-order phase transition at lower temperature. This behavior suggests the presence of a tricritical point close to this composition. For larger amounts of interstitial Fe, $y = 0.15$, once again a single phase transition is observed at 63 K in the heat capacity measurements. However, this phase transition is a continuous\cite{Ros2011}  ($\lambda-$like in specific heat)~transition from tetragonal paramagnetic to orthorhombic incommensurate antiferromagnetic phase. \cite{Bao2009,Rodriguez 2011} The microscopic mechanisms driving these phase transitions are not yet well understood.\\

\begin{figure}[t]
\centering 
\includegraphics[width=8.5 cm,clip]{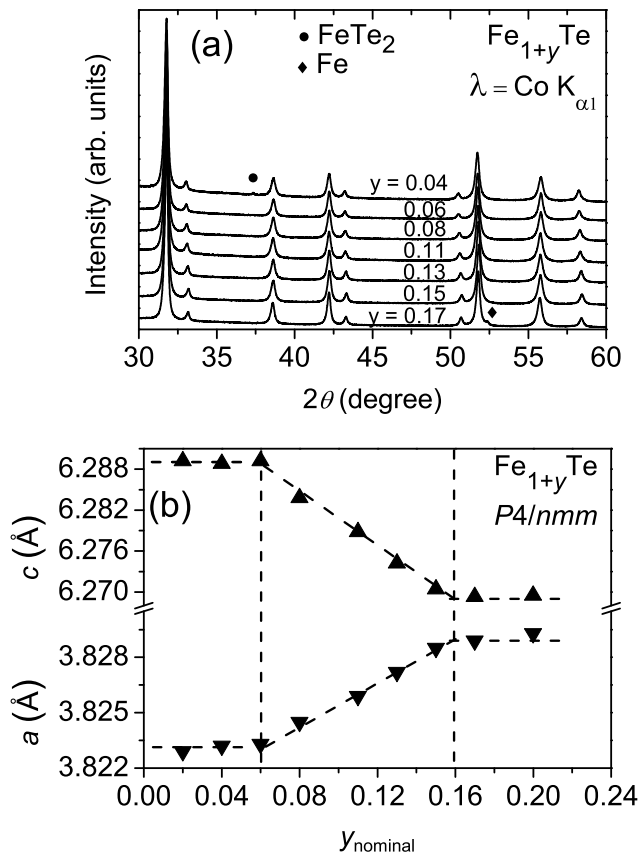}
\caption{X-ray diffraction diagram of samples with nominal composition Fe$_{1+y}$Te for $y = 0.04-0.17$, tetragonal Fe$_{1+y}$Te as the main phase at room temperature. (Impurity phases; FeTe$_{2}$ marked by \textbullet~and elemental Fe by $\blacklozenge$) (b) Lattice parameters at room temperature in dependence of the nominal composition Fe$_{1+y}$Te. The error bars here are smaller than the size of the symbols.}
\label{fig1}
\end{figure}
A strong influence of excess Fe on the magnetic and crystallographic properties of  Fe$_{1+y}$Te ($y = $ 0.076, 0.141, and 0.165 ) was first reported by Bao $et~ al.$ based on neutron diffraction experiments. \cite{Bao2009} Following this report, several other groups made similar observations. \cite {Rodriguez 2011, Zaliznyak2012, Ros2011, Mizuguchi2012}  However, due to extreme sensitivity of the physical properties of Fe$_{1+y}$Te  to the amount of $y$, it is often difficult to compare the results of independent  measurements.  Furthermore, Rodriguez $et~ al.$ reported a phase diagram\cite{Rodriguez 2011} of  Fe$_{1+y}$Te for Fe:Te in the nominal range 1.04$-$1.18:1, while a report by Mizuguchi $et~al.$  extended the phase diagram\cite{Mizuguchi2012} up to 1.3:1. These  results suggest an ambiguity in the homogeneity range of the room temperature tetragonal phase of  Fe$_{1+y}$Te. Therefore, our goal here is to establish the homogeneity range based on careful x-ray diffraction experiments and physical property measurements on chemically well characterized samples. In our previous study,\cite{Ros2011} we presented a  tentative phase diagram of  Fe$_{1+y}$Te,  which is incomplete around the composition $y=0.11$. In the case of Fe$_{1.13}$Te, we reported  two thermodynamic anomalies, and assigned the  phase transition  at lower temperature $T_{s} = 46 $~K to the structural transformation.  \cite{Ros2011} However, a recent report  \cite{Mizuguchi2012}~on the same nominal composition by Mizuguchi $et~al.$ shows a two-step structural phase transition, from tetragonal$-$orthorhombic followed by orthorhombic$-$monoclinic structure upon cooling. Further, the neutron diffraction  data on  Fe$_{1.10}$Te with similar thermodynamic properties like our Fe$_{1.13}$Te indicated a structural anomaly at 63 K followed by  a long-range magnetic order at 57.5~K.  These different results may also be related to subtle differences in the Fe content. \cite{Zaliznyak2012} Here, we focus on the detailed analysis of the powder diffraction patterns and the temperature dependence of the peak-width in Fe$_{1+y}$Te within the range 0.11 $\leq y \leq $ 0.15 to understand which phases are involved close to the tricritical point in the Fe$_{1+y}$Te phase diagram.  We aim to fill-in  the gaps as well as revise the phase diagram  to gain a clearer picture of the interplay between structure and magnetism in these compounds.
\section{Experimental}
Polycrystalline Fe$_{1+y}$Te samples were synthesized utilizing the solid-state reaction method as described in Ref. 16 with different amounts of excess iron in the range 0.02 $\le y \le$ 0.20. Prepared samples were investigated by x-ray powder diffraction (XRD) using Co K$_{\alpha1}$ radiation ($\lambda = 1.788965$ \AA). The lattice parameters of samples were calculated with LaB$_{6}$ as an internal standard in the x-ray powder diffraction experiments. As the amount of excess iron is extremely important for the physical properties of Fe$_{1+y}$Te, the synthesized phase-pure samples were characterized by wavelength dispersive x-ray (WDX) analysis and the inductively coupled plasma (ICP) method to determine the amount of Fe. 
The specific heat $C_{p}(T)$ and electrical resistivity $\rho(T)$ were measured employing a Quantum Design  physical property measurement system (PPMS). The magnetic susceptibility $\chi (T)$ was obtained by means of a SQUID magnetometer. The powders of polycrystalline materials for synchrotron measurements were ground from exactly the same pieces that were used for heat capacity and magnetic susceptibility measurements, in order  to correlate the structural phase transitions with the physical properties at a given composition. The diffraction data were collected on the high resolution powder diffraction beamline ID31 ($\lambda = 0.43046 $~\AA) at the ESRF, Grenoble, using a special He-flow cryostat adapted to the diffraction setup environment. Lattice parameter determination and structure refinements were performed by the least-squares method using JANA2006. \cite{jana2006} In Rietveld refinement procedures, anisotropic strain broadening and the March-Dollase approach for describing the preferred orientation were applied. \cite{March1932,Dollase1986}
\begin{figure}[tb]
\centering 
\includegraphics[width=8.5 cm,clip]{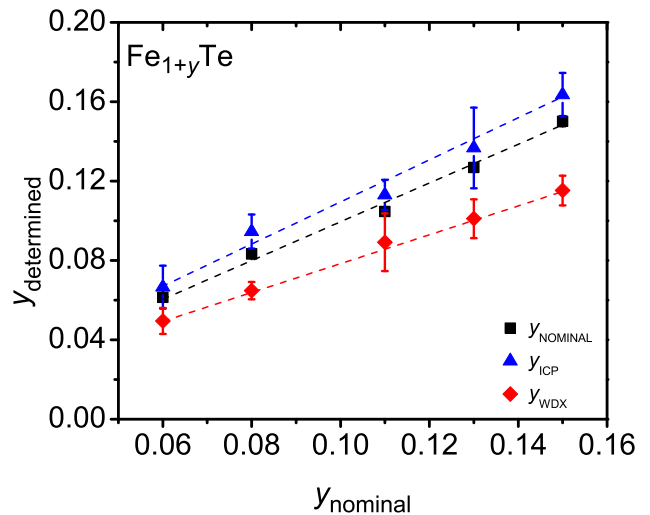}
\caption{Determined compositions of Fe$_{1+y}$Te with wavelength dispersive x-ray (WDX) analysis and chemical analysis by an inductively coupled plasma method (ICP). In calculating the standard deviation of the nominal compositions mass loss after reaction was accounted for assuming that all loss is caused by tellurium evaporation.  However, this error bar is smaller than the symbol size. }
\label{fig2}
\end{figure}

\section {Results and discussion}

 The XRD patterns of Fe$_{1+y}$Te ($ y = $ 0.04, 0.06, 0.08, 0.11, 0.13, 0.15, and 0.17) samples at ambient temperature are presented in Fig. 1 (a).  FeTe$_{2}$ and Fe impurities were observed in XRD patterns for compositions $y <0.06 $~and $y ~>$  0.15, respectively.  Previously reported excess amount of iron in tetragonal Fe$_{1+y}$Te ranged from 0 to 30 \%. \cite{Rodriguez 2011, Mizuguchi2012,Ciba1955,Ipser1974, Okamoto1990} According to our x-ray diffraction study and lattice parameters represented in Figs. 1(a) and (b), the homogeneity range of tetragonal Fe$_{1+y}$Te is clearly smaller than those given in these previous reports. In Fig 2, the experimentally determined composition by WDX and ICP spectroscopic method are compared to the nominal composition.
While the amount of Fe as obtained by the ICP method is systematically 1$-$2 \% higher than the nominal composition, WDX analysis gives an amount of iron that is typically 1$-$3 \% lower. The compositions obtained from WDX and chemical analysis overlap with the nominal composition within three standard deviations, 3$\sigma$. 
\begin{figure}[t!]
\centering 
\includegraphics[width=8.5 cm,clip]{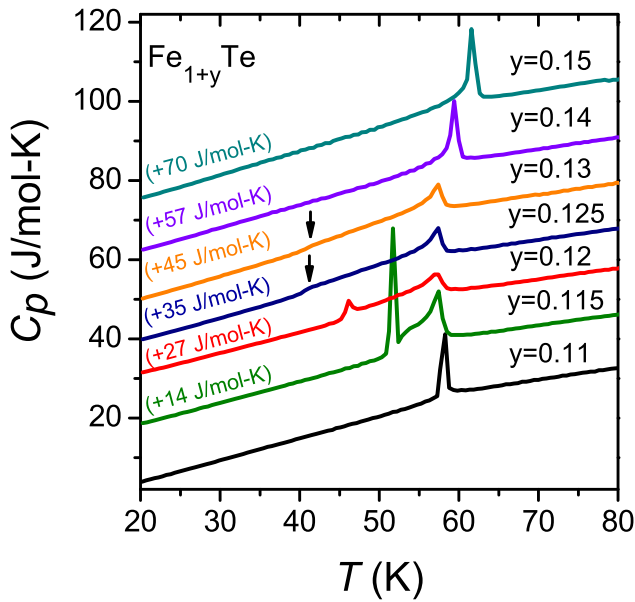}
\caption{Specific heat of Fe$_{1+y}$Te for $y = 0.11-0.15$. The $C_{p} (T)$ data for $y = 0.115-0.15$ are shifted by the amounts given for each curve for clarity. Arrows show the disappearing first-order phase transition upon increasing Fe composition.}
\label{fig3}
\end{figure}
\begin{figure}[t!]
\centering 
\includegraphics[width=8.5 cm,clip]{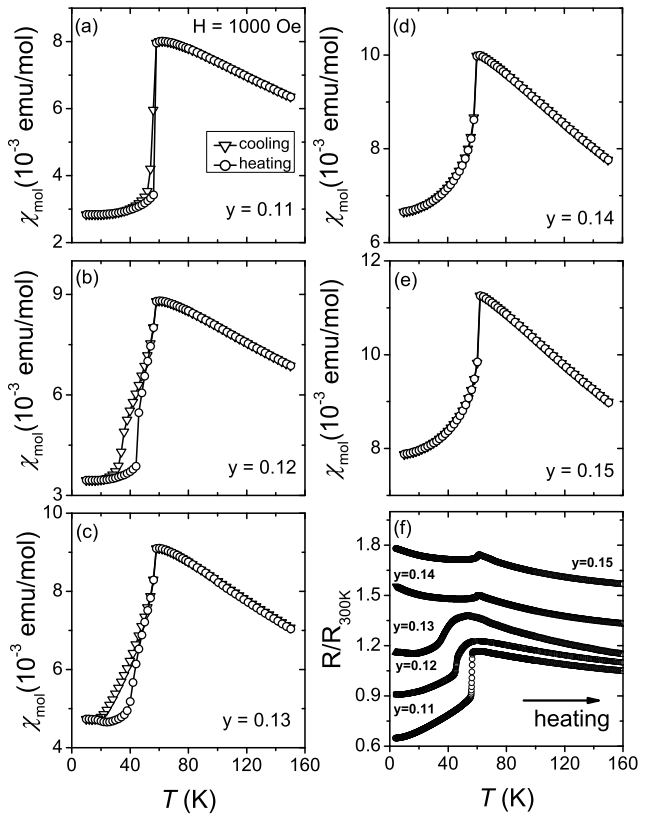}
\caption{Magnetic susceptibility of Fe$_{1+y}$Te for $y = 0.11-0.15$ (a-e) and normalized resistance (R/R$_{300 \mathrm{K}}$) during heating cycle (f). The magnetic susceptibility was measured in a field of 0.1 T. The R/R$_{300 \mathrm{K}}$ data for $y = 0.14$~ and 0.15 are multiplied by a factor of 1.25 and 1.5, respectively, for better visibility.}
\label{fig4}
\end{figure}

The temperature dependence of the specific heat of Fe$_{1+y}$Te for $y = 0.11-0.15$ is presented in Fig. 3. For y = 0.11, a peak corresponding to a simultaneous first-order  magnetic and structural phase transition at $\approx 58$~ K is observed. With minute increase in the Fe composition, however, two phase transitions can be distinguished. Already for y = 0.115 these two transitions are well separated. For the composition Fe$_{1.12}$Te, the  $\lambda-$like second order phase transition at 57 K is followed by a first-order phase transition at lower temperature, 46 K, as reported previously for a single crystal with nominal composition Fe$_{1.13}$Te. \cite{Ros2011} With increasing Fe-content, the first-order phase transition at lower temperature disappears and for $y$ = 0.14 only one transition is detected around 59 K with the characteristics of a continuous phase transition. The corresponding transition for y = 0.15 is found at a slightly increased temperature of 63 K. 

In order to compare the crystallographic phase transitions of Fe$_{1+y}$Te compositions to their magnetic and electrical properties, we performed magnetic susceptibility $\chi (T)$ and resistivity $\rho (T)$ measurements. Figs. 4(a)$-$4(e) display the temperature dependence of magnetic susceptibility measured under magnetic field of 0.1 T in field-cooling (FC) protocol for $0.11 \leq y \leq 0.15$. The magnitude of $\chi$ rises with increasing $y$ because excess Fe has a strong magnetic moment. \cite{Zhang2009} The transition temperatures obtained from specific heat and susceptibility measurements are in good agreement. The cooling and warming cycles in the susceptibility measurements exhibit a small thermal hysteresis for Fe$_{1.11}$Te, which is typical for a first-order phase transition (Fig. 4a).  This thermal hysteresis in $\chi$ is broader for samples with y = 0.12 and 0.13 for which specific heat measurements indicated the presence of two consecutive phase transitions. For even higher values of $y$, $cf.$ Figs. 4(d) and 4(e), there is no thermal hysteresis in magnetic susceptibility measurement. Such behavior is in accordance with what is expected for a continuous phase transition. Fig. 4(f) presents a summary of the temperature dependence of normalized resistance (R/R$_{300\mathrm{K}}$) measured in the heating cycle. Similar thermal hysteresis as seen in $\chi $  was observed in resistivity measurements for the same compositions (not shown here).  Below the phase transition temperatures, Fe$_{1.11}$Te shows a metallic behavior, while samples with higher Fe content, y $\geq$ 0.14, display  
increasing resistivity with decreasing temperature.
\begin{figure}[t!]
\centering 
\includegraphics[width=8.5 cm,clip]{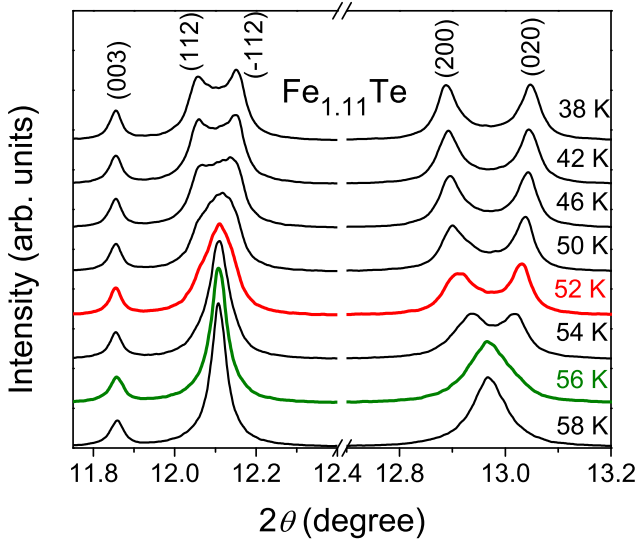}
\caption{Representative powder XRD patterns of Fe$_{1.11}$Te in the temperature regime $38-58$~ K for the (112) and (200) Bragg reflections. The green and red curves indicate an onset of orthorhombic and monoclinic distortions,  respectively.}
\label{fig5}
\end{figure}
\begin{figure}[t!]
\centering 
\includegraphics[width=8.5 cm,clip]{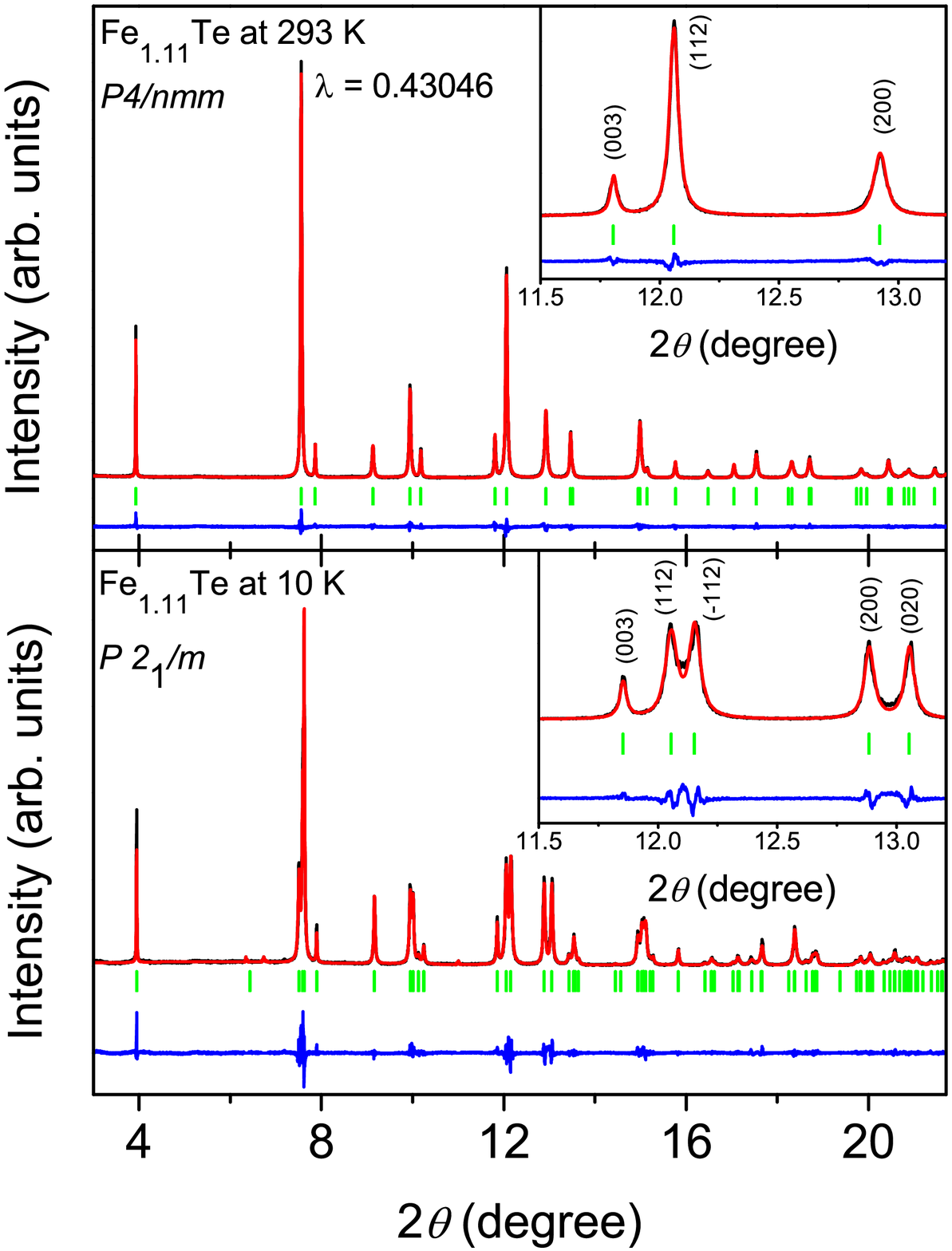}
\caption{Refined synchrotron powder x-ray diffraction patterns of Fe$_{1.11}$Te at temperatures above (293 K) and below (10 K) the phase transition. }
\label{fig6}
\end{figure}
\begin{table}[t!]
\caption{Parameters of crystal structures and refinements, atomic
positions and atomic displacement parameters $U_{\text{iso}}$
(in~10$^{-2}$ \r A$^2$) for Fe$_{1.11}$Te at room temperature and  10~K.} \label{tab1}
\begin{ruledtabular}
\begin{tabular}{lll}
Temperature  & 293~K & 10~K\\
\hline
Space group    & $P4/nmm$       &     $P2_1/m$     \\
 \hspace*{0.28cm}$a$ (\AA)     &     3.8253(3) &     3.83684(8)   \\
 \hspace*{0.28cm}$b$ (\AA)     &    = $a$       &     3.78735(8)   \\
 \hspace*{0.28cm}$c$ (\AA)     &    6.27870(6)   &     6.25409(13)    \\
\hspace*{0.28cm}$\beta$ (deg.)  & 90 &  90.668(1)  \\
\hspace*{0.28cm}$R_I/R_P$      &  0.015/0.060   &     0.015/0.089  \\
Number of reflections &  81    &      323 \\
Refined  parameters for &&\\
profile/crystal structure & 21 / 5 & 30 / 10 \\
Atomic parameters& &                                \\
\hspace*{0.28cm}Fe1     &   $2a$ ($\frac34$,$\frac14$,0) & $2e$($x$,$\frac14$,$z$) \\
        &     &  $x$ = 0.7368(4)           \\
        &    &  $z$ = $0.0004(3)$        \\
        &    $U_{\text{iso}}$ = 0.83(2)  &  $U_{\text{iso}}$= 0.68(3) \\
\hspace*{0.28cm}Fe2& $2c$ ($\frac14$,$\frac14$,$z$) &  $2e$ ($x$,$\frac14$,$z$)  \\
        &                &  $x$ = 0.277(3)            \\
        &    $z$ = 0.717(1)    &  $z$ = 0.715(2)            \\
        &       $U_{\text{iso}}$ = 0.92(2)   &  $U_{\text{iso}}$ = 1.1(2) \\
Occupancy & 0.108 (1)&0.108(0)\\				
\hspace*{0.28cm}Te      & $2c$ ($\frac14$,$\frac14$,$z$) &  $2e$ ($x$,$\frac14$,$z$)  \\
        &       &  $x$ = 0.2434(2)           \\
        &    $z$ = 0.28207(5)           &  $z$ = 0.28269(7)          \\
        &         $U_{\text{iso}}$ = 0.94(1)  &  $U_{\text{iso}}$ = 0.75(1)
\end{tabular}
\end{ruledtabular}
\end{table}
\begin{figure}[t!]
\centering 
\includegraphics[width=8.5 cm,clip]{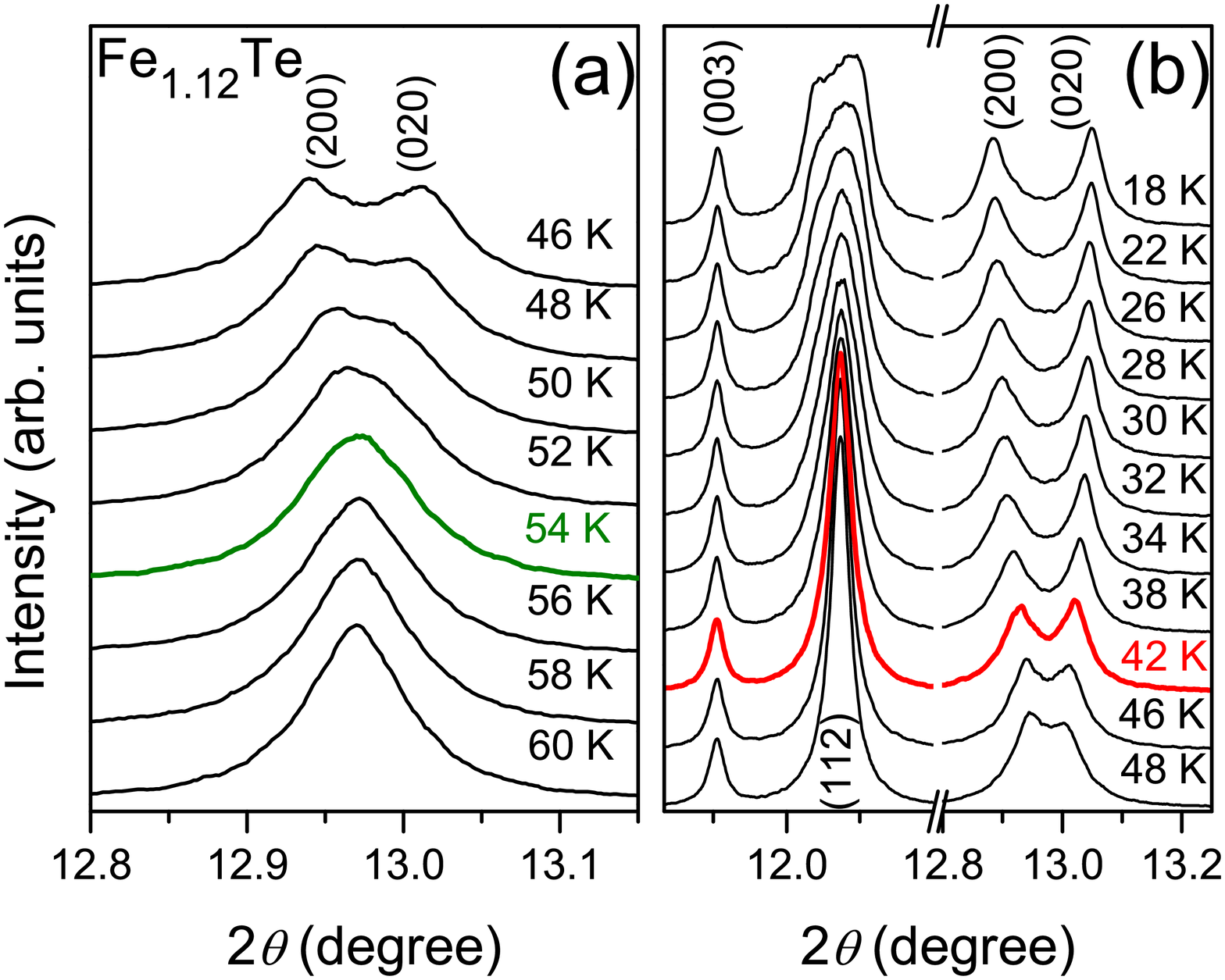}
\caption{Representative powder XRD patterns of Fe$_{1.12}$Te in the temperature regime $18-
60$ K. (a) The region of the (200) reflection between $46-60$ K. (b) The combined region of (112) and (200) reflections between 18 and 48 K.  The green and red curves indicate the onset of orthorhombic and monoclinic distortions,  respectively.}
\label{fig7}
\end{figure}
\begin{figure}[t!]
\centering 
\includegraphics[width=8.5 cm,clip]{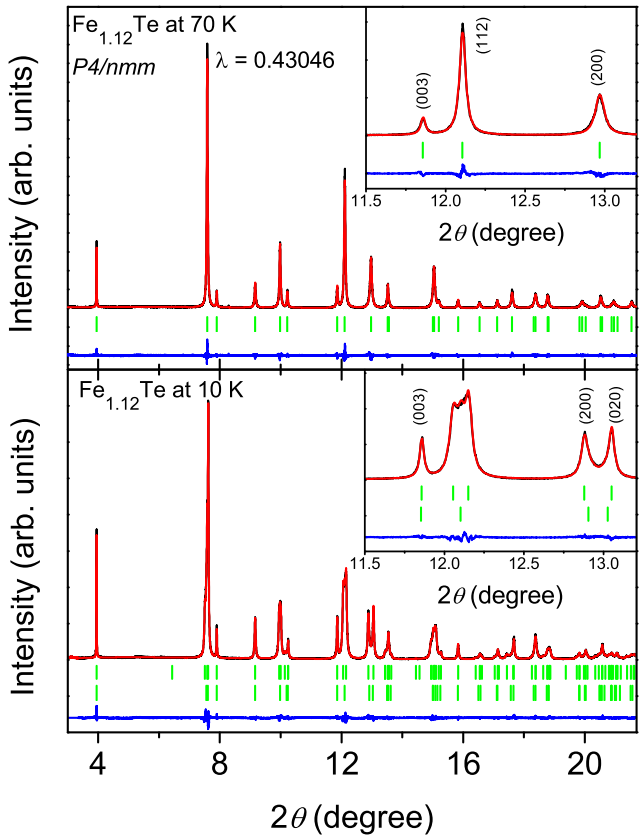}
\caption{Refined synchrotron powder x-ray diffraction patterns of Fe$_{1.12}$Te at temperatures above (70 K) and below (10 K) the phase transition. At 10 K, the upper and lower Bragg reflections represent monoclinic and orthorhombic structures, respectively.  }
\label{fig8}
\end{figure}

To correlate the physical properties with the crystal structures, we performed high-resolution synchrotron x-ray diffraction of the polycrystalline samples  from 70 to 10~K with 2~K temperature intervals. A complete structure refinement was conducted for all studied compositions. Fig. 5 represents the selected region of XRD pattern for the (112) and (200) Bragg reflections of Fe$_{1.11}$Te in the temperature regime $38-58$ K during the cooling cycle. The peak splitting of both (200) and (112) Bragg reflections is characteristic of the monoclinic ($P2_{1}/m$) phase transition in the Fe$_{1+y}$Te system. In Fig. 5, a broadening of the (200) reflection can be seen at 56 K, while the peak splits into (200) and (020) at 54 K.  A broadening of the (112) reflection is visible at 52 K and the splitting into (112) and (-112) becomes more pronounced at lower temperatures.  A full-profile refinement of powder XRD data of  Fe$_{1.11}$Te at room temperature and 10 K are given in Fig. 6. According to the Rietveld refinement, the composition is determined as Fe$_{1.108(1)}$Te, which is consistent with the nominal composition. The refined data confirm the temperature-induced transformation from tetragonal ( $P4/nmm$ at 293 K) to the monoclinic phase ($P2_{1}/m$ at 10 K) at low temperature. Refined parameters of the crystal structures are represented in Table I.  Note that there is no indication for any presence of an orthorhombic phase in Fe$_{1.11}$Te at 10~K.

In the case of Fe$_{1.12}$Te with two distinct phase transitions, the broadening of the (200) reflection starts at around 54 K and the splitting is visible at 50 K (Fig. 7(a)). However, for the (112) peak, no apparent change of the peak shape was observed down to 42 K, see Fig. 7(b). Below 42 K, the (112) peak starts broadening but no clear splitting is observed even at the base temperature, 10 K, in contrast to Fe$_{1.11}$Te. Our observations confirm that Fe$_{1.12}$Te consists of a mixture of orthorhombic ($Pmmn$) and monoclinic ($P2_{1}/m$) phases at low temperature, as reported by Rodriguez $et~al.$ \cite{Rodriguez 2011} From the results of specific heat and synchrotron XRD measurements, the $\lambda-$like second order phase transition at 57 K is associated with the structural phase transition from tetragonal to orthorhombic symmetry, while the first-order phase transition observed in the specific heat measurements at 46 K corresponds to an incomplete orthorhombic to monoclinic phase transition. The latter phase transition in Fe$_{1.12}$Te is sluggish because of a strong competition between orthorhombic and monoclinic phases.

The powder x-ray diffraction patterns of Fe$_{1.12}$Te at several temperatures were investigated by Rietveld refinement to determine the crystal structure at different temperatures.   At 70 K, the XRD pattern can be refined as a single tetragonal phase. However, at 10 K, the XRD pattern of Fe$_{1.12}$Te can only be fitted reasonably as a mixture of orthorhombic and monoclinic phases (Fig. 8).  The relative fractions of the phases (in wt.\%) ~ are ~ 65\% monoclinic ($P2_{1}/m$) and ~ 35 \% orthorhombic ($Pmmn$)~at 10 K. According to Mizuguchi $et~al.$, the estimated population of orthorhombic phase at 5 K is $20-30\% $ which is close to our results at 10 K.\cite{Mizuguchi2012} The details of the refinement of Fe$_{1.12}$Te  are compiled in Table~II. 

%
%
%
\begin{table*}[t]
\caption{Parameters of crystal structures and refinements, atomic
positions and atomic displacement parameters $U_{\text{iso}}$
(in~10$^{-2}$ \r A$^2$) for Fe$_{1.12}$Te  in the tetragonal phase at 70~K and in
the mixed phase at 10~K.} \label{tab2}
\begin{ruledtabular}
\begin{tabular}{llll}
Temperature  & 70~K & 10~K&10~K\\
\hline
Space group    & $P4/nmm$       &     $P2_1/m$  & $Pmmn$    \\
 \hspace*{0.28cm}$a$ (\AA)     &     3.81200(5) &     3.83845(4)& 3.82971(1)   \\
 \hspace*{0.28cm}$b$ (\AA)     &    = $a$       &     3.78807(3)&3.79463 (1)   \\
 \hspace*{0.28cm}$c$ (\AA)     &    6.25119(9)   &     6.25193(5)& 6.2521 (1)    \\
\hspace*{0.28cm}$\beta$ (deg.)  & 90 &  90.649(1)&90  \\
\hspace*{0.28cm}$R_I/R_P$      &  0.013/0.084   &     0.013/0.054 & 0.008/0.054  \\
Number of reflections &  80    &      232 &133 \\
Refined  parameters  for &&\\
profile/crystal structure & 22 / 5 & 36 / 11 & 36 / 11 \\
Atomic parameters& &                                \\
\hspace*{0.28cm}Fe1     &   $2a$ ($\frac34$,$\frac14$,0) & $2e$($x$,$\frac14$,$z$)&  $2b$ ($\frac34$,$\frac14$,$z$)\\
        &      &  $x$ = 0.7378(3)&           \\
        &                                &  $z$ = $0.0019(3)$   &   $z$ = 0.0042 (7)   \\
        &       $U_{\text{iso}}$ = 0.3418(2)  &  $U_{\text{iso}}$= 0.2(0)&$U_{\text{iso}}$= 0.2(0) \\
\hspace*{0.28cm}Fe2$^a$ & $2c$ ($\frac14$,$\frac14$,$z$) &  $2e$ ($x$,$\frac14$,$z$)&$2a$ ($\frac14$,$\frac14$,$z$)  \\
        &                &  $x$ = 0.258(3)&            \\
        & $z$ = 0.720(1)  &  $z$ = 0.714(2)&  $z$ =0.745(3)            \\
        &   $U_{\text{iso}}$ = 0.3(0)&  $U_{\text{iso}}$ = 0.2(0)&  $U_{\text{iso}}$ = 0.2(0) \\
\hspace*{0.28cm}Te      & $2c$ ($\frac14$,$\frac14$,$z$) &  $2e$ ($x$,$\frac14$,$z$)  \\
        &               &  $x$ = 0.2432(2)     &      \\
        &  $z$ = 0.28319(7)    &  $z$ = 0.2842(1)   &  $z$ = 0.2805(3)        \\
        &      $U_{\text{iso}}$ = 0.3726(1)                            &  $U_{\text{iso}}$ = 0.2(0)  &  $U_{\text{iso}}$ = 0.2(0)
\footnotetext{For the refinement involving two phases, the occupancy of the Fe2 site was fixed at $y = 0.12$.}
\end{tabular}
\end{ruledtabular}
\end{table*}
%
%

At higher Fe content, y = 0.14, the broadening of the (200) peak appears at 54 K and visible splitting is monitored at around 50 K (Fig. 9).  As expected for an orthorhombic symmetry, the (112) peak does not exhibit broadening or splitting even at lowest measured temperature. Refined synchrotron powder x-ray diffraction patterns of Fe$_{1.14}$Te at room temperature and 10~K are given in Fig. 10. At 10 K, the XRD pattern of Fe$_{1.14}$Te can be refined assuming a  pure orthorhombic phase. Refined parameters of crystal structures at 293 and 10 K are listed in Table~III. 
\begin{figure}[t!]
\centering 
\includegraphics[width=8.5 cm,clip]{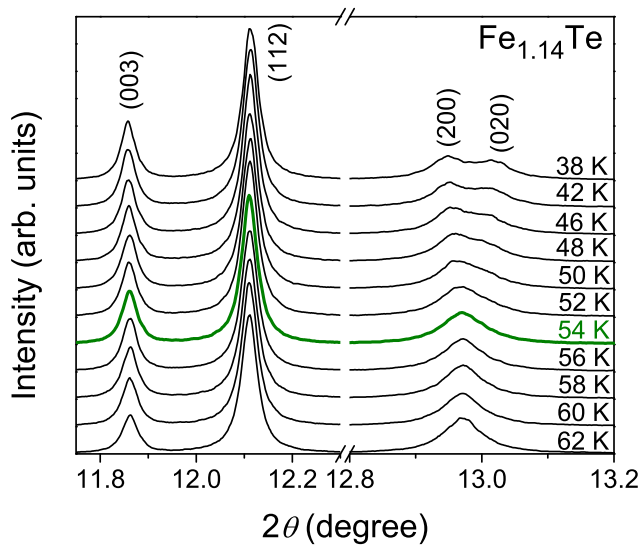}
\caption{Representative powder XRD patterns of Fe$_{1.14}$Te for the (112) and (200) Bragg reflections in the temperature regime $38-62$ K.  Broadening of the (200) peak sets in at 54 K (green line). }
\label{fig9}
\end{figure}
\begin{figure}[t]
\centering 
\includegraphics[width=8.5 cm,clip]{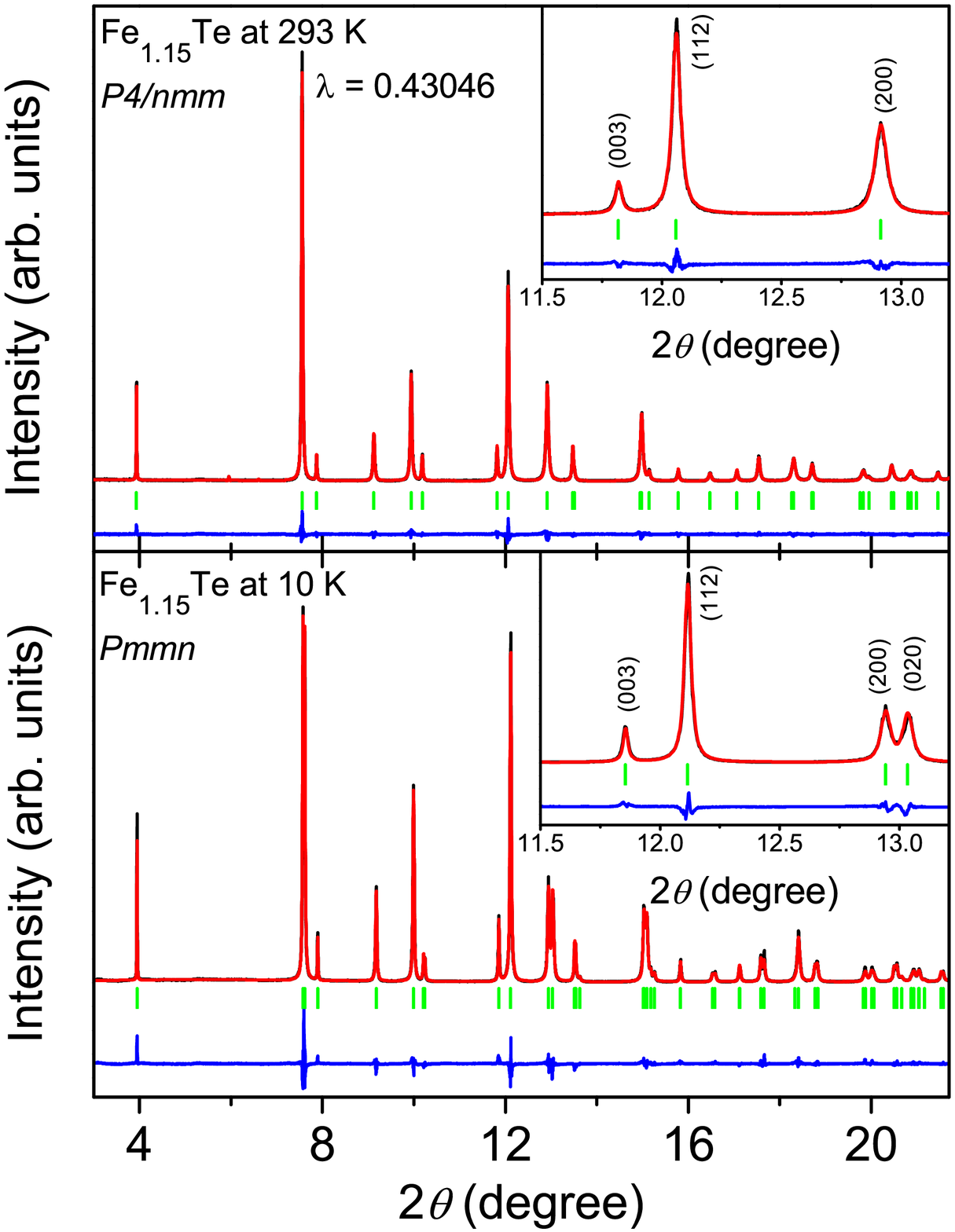}
\caption{Refined synchrotron powder x-ray diffraction patterns of Fe$_{1.15}$Te at temperatures above (293 K) and below (10 K) the phase transition. }
\label{fig10}
\end{figure}
\begin{figure}[h!]
\centering 
\includegraphics[width=8.5 cm,clip]{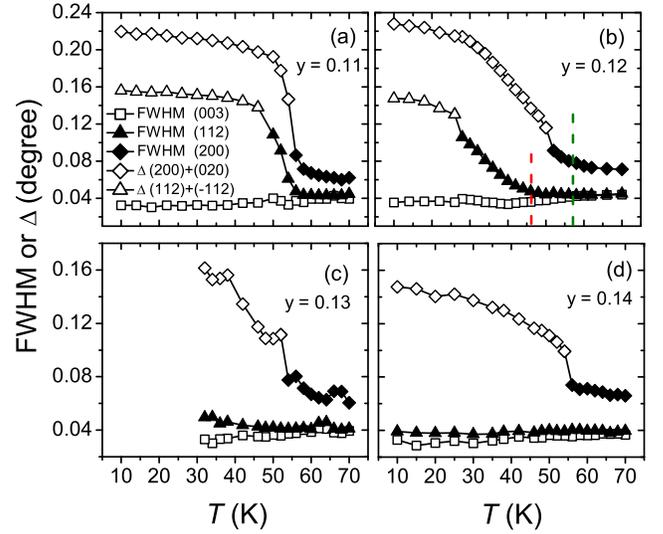}
\caption{Temperature dependence of powder x-ray diffraction peaks of Fe$_{1+y}$Te, $y = 0.11-0.14$. (a-c) The broadening of the reflections (112) and (200) demonstrate a monoclinic distortion at low temperatures, whereas in(d) constant values for (112) indicate an orthorhombic low-temperature phase. Dashed lines in (b) were drawn to mark the temperatures at which phase transitions occur in the thermodynamic measurements.  }
\label{fig11}
\end{figure}
\begin{figure}[h!]
\centering 
\includegraphics[width=7.5 cm,clip]{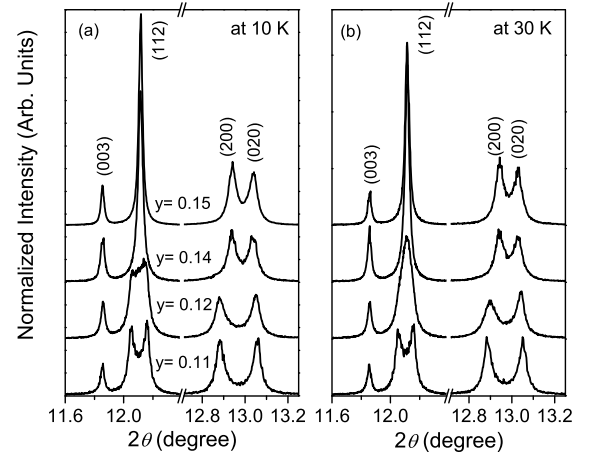}
\caption{Representative powder XRD patterns of Fe$_{1+y}$Te, $y = 0.11-0.15$ at 10 K (a) and 30 K (b).  For $y =$ 0.11, both (200) and (112) peaks are clearly split at low temperatures confirming the monoclinic structure.  The partially split (112) peak in the case of $y =$ 0.12 sample shows that the compound is a mixture of  orthorombic and monoclinic phases at 10 K.  The unsplit (112) peak at 10 K for $y =$ 0.14 and $y =$ 0.15 samples confirm pure orthorhombic phase at 10 K. }
\label{fig12}
\end{figure}
\begin{figure}[h!]
\centering 
\includegraphics[width=9.07 cm,clip]{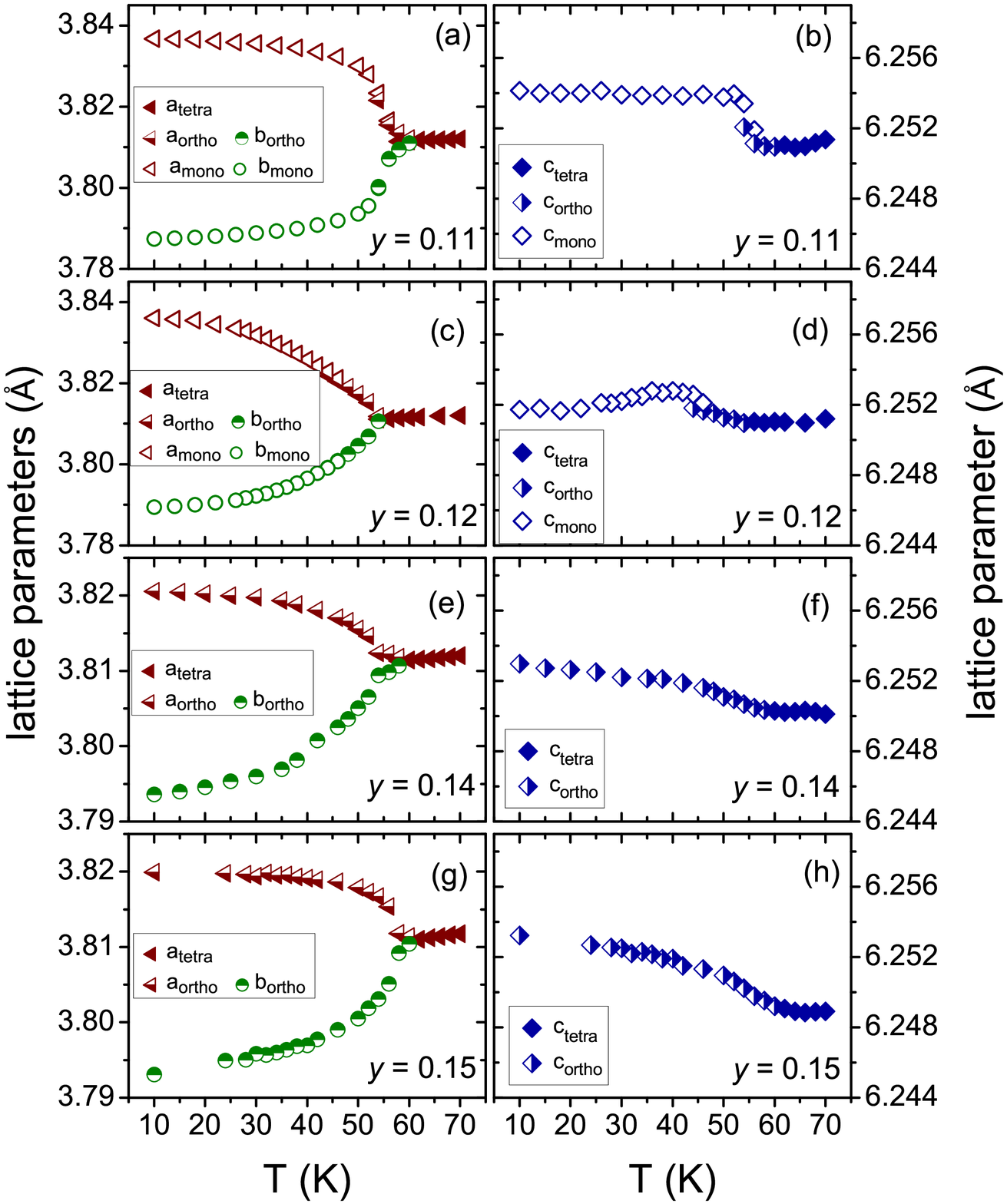}
\caption{Temperature dependence of lattice parameters $a, b$ and $c$ at various compositions. (a)-(d) Transition from tetragonal to monoclinic symmetry (with intermediate orthorhombic phase, represented by half full symbols)for y = 0.11 and 0.12 . (e)-(g)  Orthorhombic phase transition for y = 0.14 and 0.15. Below 46 K Fe$_{1.12}$Te  consists of a mixture of orthorhombic and monoclinic phases. In (c) and (d), the lattice parameters below 46 K were evaluated assuming a monoclinic structure exclusively. }
\label{fig13}
\end{figure}
\begin{figure}[h!]
\centering 
\includegraphics[width=8.5 cm,clip]{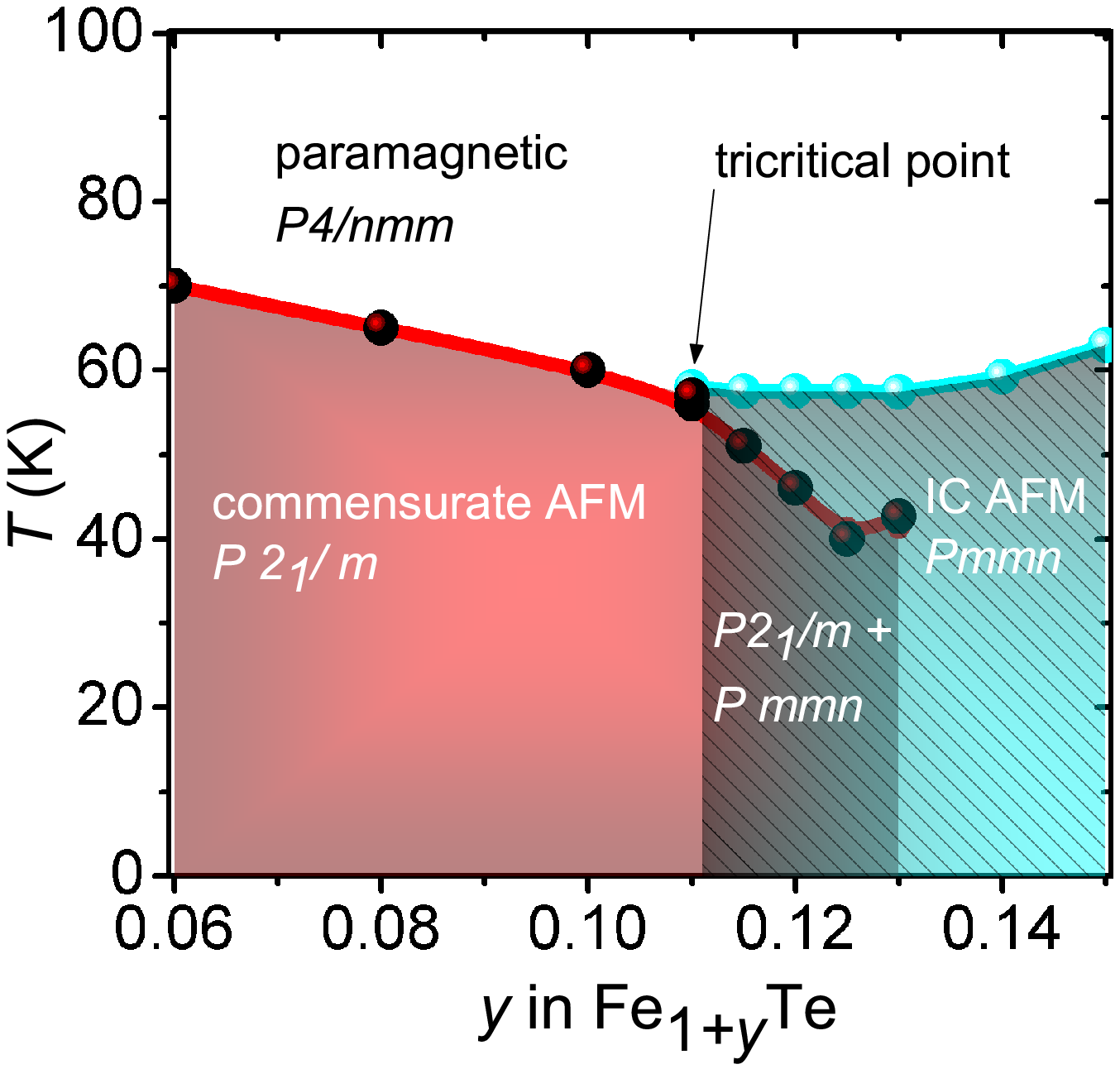}
\caption{Revised temperature-composition phase diagram of Fe$_{1+y}$Te.  AFM and IC AFM stand for antiferromagnetic and incommensurate antiferromagnetic phase, respectively.}
\label{fig14}
\end{figure}

We analyzed the full-width at half maximum (FWHM) of selected reflections below 70 K for all studied compositions in order to detect broadening and/or splitting of the reflections.
The (112) and (200) reflections were selected as identification of symmetry  breaking whereas (003) was taken as a reference because its peak shape does not change across the structural transitions. Results are shown in Fig. 11. Here, $\Delta$ is given \cite{Koz2012} as the sum of the peak FWHM plus the separation of the peak maxima in case of visible splitting, i.e. a value which increases significantly upon peak splitting. In Fig. 11(a), the magnitude of FWHM of both (200) and (112) reflections in Fe$_{1.11}$Te starts to increase almost at the same temperature around 58 K. The difference of $\approx 2$ K between broadening of (200) and (112) peaks as mentioned earlier, is difficult to resolve in this analysis. However recent neutron diffraction measurements of Fe$_{1.11}$Te single crystals\cite{Oliver2013} indicate an incommensurate antiferromagnetic precursor phase before a commensurate antiferromagnetic phase sets in.  In contrast, the separation between transitions is much more pronounced for the composition Fe$_{1.12}$Te, see Fig. 11(b): The (200) reflection broadens at $\approx 57$ K whereas the value of FWHM of the (112) remains constant until 46 K. These temperatures are in conformity with the specific heat measurements. 
For Fe$_{1.13}$Te polycrystalline sample,(Fig. 11(c)) broadening in the (112) reflection is not as clear as for the previous compositions. Moreover there is a slight increase below 40 K, which coincides with the weak first-order phase transition monitored around the same temperature in specific heat. In Fig. 11(d), for $y =$ 0.14, no change in the (112) reflections is observed while broadening in (200) reflections is quite obvious because of the transition into orthorhombic symmetry. But the changes in the FWHM values of the (200) reflections for both $y=$0.14 and $y=$0.15 (not shown) compositions were observed at $3-4$~ K lower than the corresponding antiferromagnetic ordering temperature $T_{N}$. The FWHM analyses, in general,  show that the onset temperatures of the phase transitions determined by heat capacity measurements are in conformity with the results of synchrotron XRD measurements.
\begin{table}[t]
\caption{Parameters of crystal structures and refinements, atomic
positions and atomic displacement parameters $U_{\text{iso}}$
(in~10$^{-2}$ \r A$^2$) for Fe$_{1.15}$Te at room temperature and  10~K.} \label{tab1}
\begin{ruledtabular}
\begin{tabular}{lll}
Temperature  & 293~K & 10~K\\
\hline
Space group    & $P4/nmm$       &     $Pmmn$     \\
 \hspace*{0.28cm}$a$ (\AA)     &     3.82835(2) &     3.81971(3)   \\
 \hspace*{0.28cm}$b$ (\AA)     &    = $a$       &     3.79288(3)   \\
 \hspace*{0.28cm}$c$ (\AA)     &    6.27019(4)   &     6.25288(5)    \\
\hspace*{0.28cm}$\beta$ (deg.)  & 90 &  90  \\
\hspace*{0.28cm}$R_I/R_P$      &  0.022/0.067   &     0.021/0.073 \\
Number of reflections &  152    &      133 \\
Refined  parameters for &&\\
profile/crystal structure & 22 / 5 & 24/ 7 \\
Atomic parameters& &                                \\
\hspace*{0.28cm}Fe1     &   $2a$ ($\frac34$,$\frac14$,0) & $2b$($\frac34$,$\frac14$,$z$) \\
        &                                &  $z$ = $0.0020(2)$        \\
        &    $U_{\text{iso}}$ = 0.94(1)                            &  $U_{\text{iso}}$= 0.60(2) \\
\hspace*{0.28cm}Fe2& $2c$ ($\frac14$,$\frac14$,$z$) &  $2a$ ($\frac14$,$\frac14$,$z$)  \\
        & $z$ = 0.7175(5)                &  $z$ = 0.7159(8)            \\
        &  $U_{\text{iso}}$ = 0.80(7)                                  &  $U_{\text{iso}}$ = 0.5(1) \\
Occupancy & 0.152 (1)&0.152\\				
\hspace*{0.28cm}Te      & $2c$ ($\frac14$,$\frac14$,$z$) &  $2a$ ($\frac14$,$\frac14$,$z$)  \\
        & $z$ = 0.28400(3)               &  $z$ = 0.28490(5)           \\
        & $U_{\text{iso}}$ = 1.09(1)     &   $U_{\text{iso}}$ = 0.54(1)         \\         
\end{tabular}
\end{ruledtabular}
\end{table}

In Figs. 12(a) and (b), the selected region of XRD patterns for (112) and (200) Bragg reflections of Fe$_{1+y}$Te, $y = 0.11-0.15$, are given at 10 K and 30 K to summarize the low-temperature behaviors of different compositions. Fig. 12(b) exhibits that the samples Fe$_{1+y}$Te with $y \geq 0.14$  are clearly orthorhombic while Fe$_{1.11}$Te is in monoclinic phase already at 30 K. On the other hand, Fe$_{1.12}$Te at 30 K seems to be mostly in the orthorhombic phase because the peak splitting in (112) is not significant. At 10 K, the peak is broader but still there is not a clear splitting as a result of the mixture of orthorhombic and monoclinic phases. 

Our results on Fe$_{1.12}$Te are supporting the idea of a two-step evolution of the crystal structure from tetragonal via orthorhombic to monoclinic structures as suggested by Mizuguchi $et~al.$ \cite{Mizuguchi2012} In our previous report on  Fe$_{1.13}$Te  single crystals,\cite{Ros2011} only one structural phase transition was identified within the magnetically ordered phase. In any case, the present detailed investigations suggest that the low-temperature transition from orthorhombic to monoclinic phase is incomplete even at 10~K for these compositions. According to Martinelli $et~al.$ \cite{Martinelli2010} and our results, for lower Fe content, y $<$ 0.11, the phase transition from tetragonal to monoclinic does not need an intermediate phase (orthorhombic) formation.  But in the vicinity of a tricritical point on the right-hand side, the intermediate orthorhombic phase partially transforms towards monoclinic symmetry.

For a comparison of the metrical changes, the temperature dependence of the lattice parameters obtained from the refinements of several compositions $0.11 \leq y \leq 0.15 $ during the cooling cycle are summarized in Figs. 13(a)$-$13(h). The splitting of  lattice parameter $a$ at around $T_{N}$~is quite dramatic but remains almost constant throughout the monoclinic phase. In the orthorhombic phase, the difference between  lattice parameters $a$ and $b$ is significantly smaller. The difference between the first-order and second order phase transitions can be clearly seen in the $c$ parameters: For the monoclinic phase transition the increase of the $c$ parameter is sudden at around  $T_N$(Fig. 13(b)), whereas for the orthorhombic phase transition it changes smoothly (see Figs. 13(f) and (h)).  The diffraction patterns of Fe$_{1.11}$Te can be refined as either purely orthorhombic or purely monoclinic phase down to 54 K without a significant difference in the residuals and lattice parameters. In Figs. 13(a) and 13(b), the overlap of lattice parameters for both phases can be seen between 60 and 54~K. Between $46-54$~ K, the lattice parameters of Fe$_{1.12}$Te were refined as orthorhombic phase. Below 46~K the lattice parameters were calculated assuming only a monoclinic phase for simplicity. Yet, even when the diffraction pattern were refined allowing for a mixture of two phases, the lattice parameters of the monoclinic structure did not exhibit a significant difference compared to fitting a  purely monoclinic phase. In Fig. 13 (c) and (d)  however, we show the lattice parameters of only monoclinic phase for clarity.

On the basis of our results, we propose a revised temperature-composition phase diagram of Fe$_{1+y}$Te,  Fig. 14. For the lower Fe excess, $viz$, for $y < 0.11$,  the paramagnetic tetragonal phase transforms into monoclinic commensurate antiferromagnetic phase without an intermediate phase formation while $T_{N}$ decreases from 69 K to 58 K with increasing Fe amount (as suggested in Ref. 14).  A tricritical point is situated close to the composition $y \approx$ 0.11 in the phase diagram. At composition, $y = $~0.115, a two-step phase evolution is apparent.  At 10 K, for $ 0.115 \le y \le 0.13$,  the materials are composed of a mixture of monoclinic and orthorhombic phases.  The temperature difference between these transitions becomes more distinct upon increasing Fe amount.   For $y > 0.13$,  the phase transition from orthorhombic to monoclinic structure at lower temperature disappears and only single phase transition is observed. The latter is a second order phase transition from the  tetragonal paramagnetic to orthorhombic  incommensurate antiferromagnetic structure, which is in accordance with the neutron scattering experiments.\cite{Rodriguez 2011} However,  the mysterious disappearance of the first order phase transition for $y > 0.13$ requires more closer examination.  \\
\section{Conclusions}
We provide a reference data base for cross-comparing different reports on  Fe$_{1+y}$Te by conducting low-temperature synchrotron x-ray diffraction experiments, thermodynamic and resistivity measurements on a single series of chemically well-characterized samples.  Based on these data we presented a revised phase diagram for Fe$_{1+y}$Te. A closer examination suggests a
region of orthorhombic crystal symmetry for $y > 0.13$.  Further, for 0.11 $< y \leq 0.13$ the transition into orthorhombic crystal symmetry is followed by a two phase region at even lower temperature where also a monoclinic phase is found.  Along with coinciding magnetic and structural phase transitions for $y<$ 0.11, a two-step phase transition for $0.11\leq y \leq 0.13 $  was observed with both phase transitions having  magnetic and structural components. This behavior indicates a strong magneto-elastic coupling in this system.  However,  details of the microscopic couplings and the origin of this complex interplay of magnetic and structural transitions in dependence of the Fe-content is yet to be explored.
\begin{acknowledgments}
The authors  thank G. Auffermann for chemical analysis, U. Burkhardt for WDX analysis, and Walter Schnelle for electrical resistivity measurements. We acknowledge support by Andy Fitch and Yves Watier at beam-line ID31, ESRF Grenoble, during the experiments based on proposal No. HS4825. Stimulating discussions with Yu.~Grin, and L.~H.  Tjeng are gratefully acknowledged.  The project was supported by DFG SPP 1458. AAT was partly supported by the mobilitas grant MTT-77 of the ESF.\\
\end{acknowledgments}

\end{document}